\documentstyle[seceq,epsf,wrapft]{ptptex}

\newcommand{\dn}{\rm dn}
\newcommand{\cn}{\rm cn}
\newcommand{\sn}{\rm sn}
\newcommand{\nn}{\nonumber}

\newcommand{\Half}{\frac{1}{2}}
\newcommand{\E}{{\cal E}} 
\newcommand{\p}{\partial}
\newcommand{\tr}{\mathop{\rm Tr}}
\newcommand{\wt}[1]{\widetilde{#1}}

\title{String Junction from Worldsheet Gauge Theory}

\author{
Koji {\sc Hashimoto}\footnote{
Supported in part by a Grant-in-Aid for Scientific Research from
the Ministry of Education, Science and Culture of Japan(\#3160).
E-mail: hasshan@gauge.scphys.kyoto-u.ac.jp}
}
\inst{Department of Physics, Kyoto University, Kyoto 606-8502, Japan}

\recdate{February 12, 1999
}

\abst{
We study the three-pronged strings (three string junctions) from the
point of view of D-string worldsheet gauge theory. We justify
interpreting kink solutions in the 2-dimensional gauge theory as
pronged strings by examining BPS energy bounds obtained from this
theory. When the ends of the pronged string are on D3-branes, the
configuration preserves 1/4 supersymmetry with BPS conditions
including the Nahm equation. Using solutions of the Nahm equation
explicitly, we treat string junction configuration ending on the
D3-branes from the viewpoint of the D-string. In particular, when two 
separated pronged strings end on the same D3-branes, they interact
with each other through the D3-branes and the resultant trajectories
of three-pronged strings are found to be curved non-trivially.
}

\begin{document}
\maketitle

\section{Introduction}

Recent developments in string theory and M-theory have brought about
new ways of treating dynamics of supersymmetric gauge theories in
various  dimensions. One of the most important techniques explored in
this subject involves the use of brane configurations. Many
characteristic 
properties of gauge theories, such as gauge groups, flavour symmetry
groups, supersymmetries and BPS states, can be translated into a
language of branes. In some cases, results of string (or M) theory
have revealed new facts of field theory dynamics. One such example
is the ``multi-pronged string'' (``string junction'') configuration,
with which a class of new classical solutions in field theories were
discovered.\cite{ours,ours2,kawano,LY}

The multi-pronged string was first described in type IIB string theory
\cite{SHW} and found to be a BPS state preserving $1/4$
supersymmetry.\cite{DASG} \ This configuration has been vigorously
investigated in the context of string theory \cite{ZW,sen,CT,Gau} and
also M-theory.\cite{KRO} \ One of the noteworthy results is that string 
junctions play important roles in understanding the enhancement of
exceptional gauge groups when locations of D-branes coincide.\cite{ZW} 
\ Note that 5-branes can also form junctions, and a class of 5-brane 
webs (called ``brane-box configurations'') has been used to describe
chiral gauge theories.\cite{BB}

Though multi-pronged strings are expected to exist in type IIB
string theory, no corresponding supergravity solution has yet been
found. Hence physics near the junction point is not understood
clearly. An interesting approach to this problem of realization of the 
multi-pronged string is making use of the low-energy effective field
theory on the D-brane described by D-brane actions. Some
configurations of D-branes have been obtained in this effective field
theory,\cite{CM,Gibb,Hashi} 
where a string stuck to D3-branes was considered. For the case of
multi-pronged string, it is found in Ref.\ \citen{BER} that 
there is a close relation between $1/4$ BPS states in $4$-dimensional
gauge theories (D3-brane worldvolume gauge theories) and the
multi-pronged string configurations in string theory. From these
background developments, actual classical solutions of $1/4$ BPS
states in $4$-dimensional super Yang-Mills theory were recently
constructed in Refs.\ \citen{ours} -- \citen{LY}. Plotting the
configuration, it 
was found \cite{ours,ours2,kawano} that the component strings forming
the multi-pronged string bend non-trivially.

In this paper, we analyze configurations of multi-pronged strings
from the D-string worldsheet point of view. The low energy effective field
theory on the D-string worldsheet is a $2$-dimensional supersymmetric
gauge theory. A multi-pronged string configuration in this effective
theory is studied in Ref.\ \citen{DASG}, where infinitely long
component strings are considered. And these strings are found to be
straight so that the type IIB $SL(2,Z)$ symmetry is favored. However,
if the ends of 
the pronged string are on D3-branes, it is expected that the
multi-pronged string and D3-branes interact with each other and that
the 
configuration becomes non-trivial. This situation is very interesting, 
since there are corresponding configurations on the side of the
D3-brane effective field theory. D3-branes are incorporated
into the D-string worldsheet gauge theory in Ref.\ \citen{Dia} as a
concrete realization of Nahm equation.\cite{nahm} \ Thus we
combine these ideas and investigate the shape of the multi-pronged
string terminating at the D3-branes. 

The organization of this paper is as follows. In section \ref{2}, we
review the D-string worldsheet approach 
to the multi-pronged string given in Ref.\ \citen{DASG} and show that
the BPS energy bound of the $2$-dimensional system is consistent with
the energy given by the string picture. In section \ref{3}, D3-branes
are introduced. We examine the preserved supersymmetry, and bending
configurations of the multi-pronged strings are presented. Section
\ref{4} is devoted to summary and discussions. In the appendix, our
solution is shown to satisfy the equations of motion of the
non-Abelian Born-Infeld action proposed in Ref.\ \citen{Tseytlin}.

\section{Three-pronged string with no D3-brane}
\label{2}

Dynamics of D$p$-branes is described well by a ($p\!+\!1$)-dimensional
gauge theory, which is obtained by dimensional reduction of the
$10$-dimensional supersymmetric Yang-Mills theory.\cite{Wit} \ Thus
the relevant $2$-dimensional action is\footnote{The 
  notation of this paper follows that used in Ref.\ \citen{Nil}.} 
\begin{eqnarray}
  \label{action}
  S=T_{\rm D1}\int \! d^2x\;\tr
  \left[
    -\frac14 F_{\mu\nu}F^{\mu\nu}+\Half D_\mu \xi^I D^\mu \xi^I 
    +\frac14 [\xi^I,\xi^J]^2
  \right] + S_{\rm f}.
\end{eqnarray}
Here $T_{\rm D1}$ is the tension of the D-string. The D-string extends
in the direction (01), and the eight scalar fields $\xi^I$
($I=2,\cdots,9$) contained in the above action represent the
transverse fluctuation of the D-string. This system possesses $(8, 8)$
supersymmetry (we omit the fermion terms $S_{\rm f}$
hereafter for convenience). From the action (\ref{action}), we have
the following expression for the energy of this system:
\begin{eqnarray}
  \label{energy1}
  U=T_{\rm D1}\int\! dx \;\Half\tr
  \left[
    \E^2+(D_0 \xi^I)^2 + (D_1 \xi^I)^2 -\Half [\xi^I,\xi^J]^2
  \right].
\end{eqnarray}
The coordinate $x\equiv x^1$ denotes the spatial direction of the
worldsheet, and $\E$ is the electric field.  With this energy formula,
we shall study the BPS nature of the 2-dimensional system in section
\ref{2} and \ref{3}.

\subsection{Kink solution}

First we review the considerations given in
Ref.\ \citen{DASG},\footnote{See also Ref.\ \citen{Gau}.} \
where a three-pronged string is realized as a configuration of a
worldsheet effective gauge field theory on a single D-string. When one
deals with a single D-string, the scalar and the gauge fields are
Abelian, and hence there is no potential term. All the scalar fields
decouple from each other. We turn on only one scalar field $S
\equiv \xi^2$. Then the $(x, S(x))$-plane (which is equivalent to the
$(\xi^1,\xi^2)$-plane) is interpreted as a $2$-dimensional plane on
which the pronged string lies. 

In the picture of IIB superstring theory, the configuration of the
string junction preserves eight of the original 32 supercharges. By
the existence of the D-string on which we carry out computations,
the supersymmetry is broken to half of the original supersymmetries,
i.e., 16 supercharges in 2 dimensions are preserved at this
stage. Therefore the junction configuration observed from the D-string
standpoint is expected to preserve half of the 16
supercharges. The supersymmetry transformation of the gaugino is
\begin{eqnarray}
\label{gaugino}
\lefteqn{
  \delta\lambda_{\alpha i}=2
  \left[
    (\sigma^0\bar{\sigma}^1-\sigma^1\bar{\sigma}^0)_\alpha^{\;\beta}
    \zeta_{\beta i} \E
\right.}
\nn\\
&&\qquad\qquad\qquad
+
\left.
    (\sigma^0\bar{\sigma}^3-\sigma^3\bar{\sigma}^0)_\alpha^{\;\beta}
    \zeta_{\beta i} D_0S
    +
    (\sigma^1\bar{\sigma}^3-\sigma^3\bar{\sigma}^1)_\alpha^{\;\beta}
    \zeta_{\beta i} D_1S
  \right],
\end{eqnarray}
where $i$ denotes the index of $SU(4)$ subgroup of R-symmetry, and
$\alpha$ is a spinor index. With the BPS condition
\begin{eqnarray}
\label{BPSS}
  \E\pm D_1S=0,\qquad D_0S=0,
\end{eqnarray}
the transformation (\ref{gaugino}) vanishes if half of the
supersymmetry parameters $\zeta_{\alpha i}$ are set equal to
zero:\footnote{The $2$-dimensional fermion is denoted
  as $\zeta=(\zeta_-, \zeta_+)^T$.}
\begin{eqnarray}
\label{1/2}
\zeta_{(\pm) i}=0.
\end{eqnarray}
Thus half of the 16 supercharges are preserved with the BPS
conditions (\ref{BPSS})\footnote{BPS properties of the multi-pronged
  string are investigated using the worldsheet approach in Ref.\
  \citen{CT} in another way. }. Adopting the gauge $A_1=0$, a solution
of (\ref{BPSS}) is found under the requirement of time-independence of 
the configuration as\cite{DASG}
\begin{eqnarray}
\label{upper}
  A_0=\pm S.
\end{eqnarray}
This BPS condition is independent of the worldvolume dimension
considered, as seen in Ref.\ \citen{CM}.  

In the D-brane worldvolume theory, a fundamental string attached to
it and extending in the transverse direction is described by a source
of the gauge field on the D-brane, since the charge of the string
should be 
conserved at the endpoint of the string.\cite{Str} \ For the case of a
D$p$-brane with $p\geq 2$, one can regard a ``spike'' configuration
around the source as a fundamental string.\cite{CM} \ On the other
hand, this is not the case for the D-string ($p=1$), because it has
only one spatial dimension and the source does not form a spike.
The equation of motion for the electric field with a source located at
$x=x_0$, 
\begin{eqnarray}
\label{eqE}
  D_1\E=ng\delta(x-x_0), \qquad n\in Z,
\end{eqnarray}
has the following BPS solution:\footnote{We adopt the upper sign in
  Eq.\ (\ref{upper}) without lose of generality.}
\begin{eqnarray}
\label{solu}
  A_0=S=
  \left\{
  \begin{array}{lc}
      pg(x-x_0)+q,& (x<x_0)\\
      (pg-ng)(x-x_0)+q.& (x>x_0)
  \end{array}
  \right.
\end{eqnarray}
Here $g$ is a string coupling constant equal to the fundamental
electric charge, and the charges $n$ and $p$ are integral numbers
since electric flux on the D-string is quantized.\cite{CM}

\begin{figure}[tdp]
\begin{center}
\leavevmode
\epsfxsize=80mm
\epsfbox{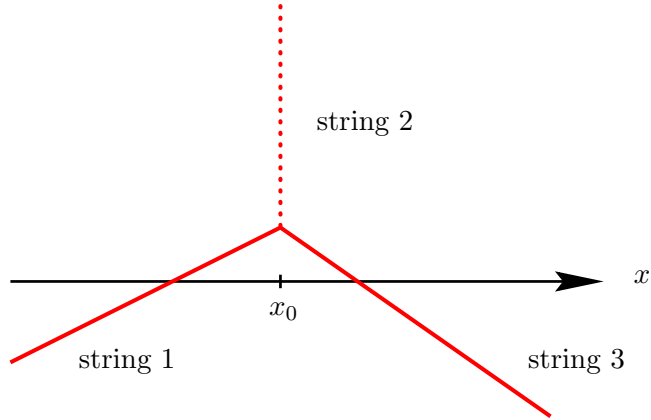}
\put(10,52){$x$}
\put(-128,40){$x_0$}
\put(-30,20){string 3}
\put(-110,110){string 2}
\put(-200,20){string 1}
\caption{The solution (\ref{solu}) representing a three-pronged
  string. The dotted line denotes the invisible fundamental string.} 
\label{fig:junct}
\end{center}
\end{figure}

As seen from the solution, the scalar field $S$ is a linear function
of $x$. There is a kink at $x=x_0$ (see Fig.\ \ref{fig:junct}), and at 
this point the electric charge is altered by $ng$. The authors of
Ref.\ \citen{DASG} interpreted this situation as follows:
{\it $n$ invisible fundamental strings\footnote{Ref.\ 
  \citen{DASG} deals with the case $n=1$. Though the state with
  $|n|\geq 2$ is marginal, our treatment in the following sections
  requires this marginal state. } are attached to the D-string, and
they form a string junction at $x=x_0$. Now the D-strings have
electric charges, thus they are interpreted as a $(-p,1)$ string
\footnote{The charge of a string is defined as the one
  following a direction oriented toward the junction point.} 
(for $x<x_0$) and a $(p\!-\!n,-1)$ string (for $x>x_0$). } 

This interpretation is justified by the fact that force balance
condition is satisfied at a junction point. Naming the regions $x<x_0$
``string 1'', $x>x_0$ ``string 3'', and the invisible fundamental
string ``string 2'', the tension of each string is read from their
charges as 
\begin{eqnarray}
  T_1=T\sqrt{p^2+\frac{1}{g^2}},\qquad
  T_2=nT,\qquad
  T_3=T\sqrt{(n-p)^2+\frac{1}{g^2}},
\end{eqnarray}
where $T$ is the tension of a single fundamental string, 
$T=g T_{\rm D1}$. 
If we assign this tension to each region of the solution, one can 
confirm that the following force balance conditions are satisfied at
the junction point $x=x_0$. One is in the horizontal direction
(parallel to $x$),
\begin{eqnarray}
  T_1\frac{1}{\sqrt{1+(pg)^2}} = T_3  \frac{1}{\sqrt{1+(pg-ng)^2}},
\end{eqnarray}
and the other is in the vertical direction (parallel to the string 2),
\begin{eqnarray}
  T_2=T_1\frac{pg}{\sqrt{1+(pg)^2}} + T_3
  \frac{ng-pg}{\sqrt{1+(pg-ng)^2}}. 
\end{eqnarray}

A tree web with more prongs can be constructed straightforwardly in
the same way. One can see that at every junction point the force
balance conditions are satisfied if the charges are conserved there.

In this way of realizingthe pronged string, the component
strings (strings 1, 2 and 3) are straight and infinitely long. Thus the 
configuration favors the $SL(2,Z)$ S-duality symmetry as discussed in 
Ref.\ \citen{DASG}. However, as we shall see in section \ref{3},
considering two three-pronged strings terminating on the same
D3-branes, they interact with each other and therefore bend
non-trivially.

\subsection{Energy of the three-pronged string}

For further justification of the above interpretation (presented in 
Ref.\ \citen{DASG}), let us investigate the energy bound of this
configuration. The energy formula (\ref{energy1}) with a single scalar
field $S$ is written as
\begin{eqnarray}
U
&&=T_{\rm D1}\int\! dx \;\Half\tr
\left[
  \E^2 + (D_0 S)^2  + (D_1S)^2
\right]\nn\\
&&=T_{\rm D1}\int\! dx \;\Half\tr
\left[
  (\E\pm D_1S)^2+(D_0 S)^2\mp2D_1 S\,\E
\right]\nn\\
&&=T_{\rm D1}\int\! dx \;\Half\tr
\left[
  (\E\pm D_1S)^2+(D_0 S)^2\mp 2\p_1 (S\E)\pm 2SD_1\E
\right].
\end{eqnarray}
Therefore using Eq.\ (\ref{eqE}), the energy is bounded by some
boundary charges and a source contribution as follows:
\begin{eqnarray}
\label{bound}
  U\geq \bigm| \E_{\rm source} + \E_{\rm boundary}\bigm|,
\end{eqnarray}
where
\begin{eqnarray}
  \E_{\rm source}=-T_{\rm D1}ng S(x=x_0)
\quad\mbox{and}\quad
  \E_{\rm boundary}=T_{\rm D1}\Bigl[\tr S\E\Bigr]_{-\infty}^{+\infty}.
\end{eqnarray}
The equality in (\ref{bound}) holds when the BPS conditions
(\ref{BPSS}) are satisfied.

The term $\E_{\rm source}$ indicates that there actually exists a
fundamental string 2, since we can express the source contribution
as 
\begin{eqnarray}
\label{Ebou}
  \E_{\rm source}
  =nT\left(-S(x\!=\!x_0)\right). 
\end{eqnarray}
The value $S(x\!=\!x_0)$ denotes the coordinate of the endpoint of the
fundamental string. Thus its length is $|\infty-S(x\!=\!x_0)|$. 
Equation (\ref{Ebou}) appropriately represents the energy of the
fundamental string with this length.

\begin{figure}[tdp]
\begin{center}
\leavevmode
\epsfxsize=80mm
\epsfbox{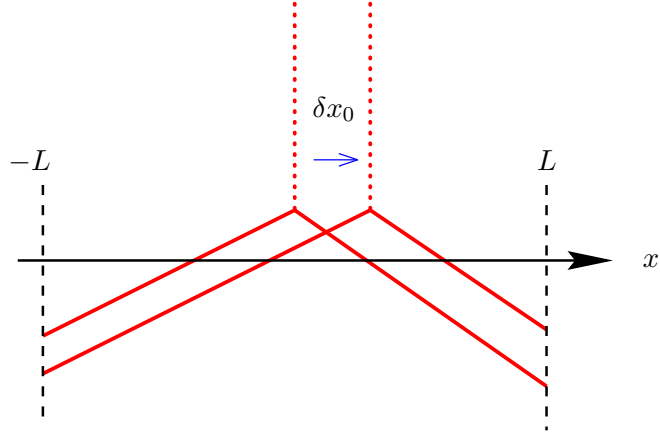}
\put(10,63){$x$}
\put(-115,120){$\delta x_0$}
\put(-30,100){$L$}
\put(-230,100){$-L$}
\caption{Moving the junction point by $\delta x_0$ in the $x$
  direction. }  
\label{fig:trans}
\end{center}
\end{figure}

On the other hand, one can see that another contribution to the
energy, $\E_{\rm boundary}$, represents strings 1 and 3. Since
$\E_{\rm boundary}$ is divergent with non-zero $n$, we restrict
the domain of $x$ to the interval $[-L,L]$ for evaluating this
term. Then for the solution (\ref{solu}), we have
\begin{eqnarray}
  \E_{\rm boundary}&&=T_{\rm D1}\Bigl[\tr
  S\E\Bigr]_{-L}^{+L}\nn\\ 
&&=T_{\rm D1}\Bigl[(pg-ng)^2(L-x_0)+(pg)^2(L+x_0)\Bigr].
\end{eqnarray}
Here we have assumed that $L$ is sufficiently large, so that $x_0\in
[-L,L]$. In order to see that the strings attached to the boundaries
$x=\pm L$ have tensions $T_1$ and $T_3$, we move the junction
point by $\delta x_0$ $(>0)$ in the $x$ direction (see Fig.\
\ref{fig:trans}). The energy changes as a result of this horizontal
translation by
\begin{eqnarray}
\label{hor}
  \delta \E_{\rm boundary}
&& \equiv 
\delta \E_{\rm boundary}(x_0+\delta x_0) - 
\delta \E_{\rm boundary}(x_0)\nn\\ 
&&=T_{\rm D1}\Bigl[(p-n)^2g^2-p^2g^2\Bigr]\delta x_0.
\end{eqnarray}
This expression is consistent with the naive argument from the string
picture; in other words, Eq.\ (\ref{hor}) can be written in the form
\begin{eqnarray}
  \left|
   \delta \E_{\rm boundary}
  \right|
=T_1 \delta l_1  +  T_3 \delta l_3 
\end{eqnarray}
with the variations of the length of string 1 and 3
\begin{eqnarray}
  \delta l_1 \equiv \delta x_0 \sqrt{1+(pg)^2},\qquad
\delta l_3\equiv - \delta x_0\sqrt{1+(ng-pg)^2}.
\end{eqnarray}

We have seen that the energy bound (\ref{bound}) appropriately
reflects the energy of a three-pronged string with no endpoint. Now
the interpretation presented in Ref.\ \citen{DASG} has been justified also
from the viewpoint of the BPS energy bound, where necessary
information regarding the energy of the three-pronged string is
encoded in Eq.\ (\ref{bound}). In particular, among three component
strings, the existence of the invisible fundamental string (string 2)
is guaranteed by the source contribution of the energy bound. In the
next section, we consider the effects of the presence of D3-branes.

\section{The Nahm equation and effects of D3-branes}
\label{3}

In this section, we take into account D3-branes in the context
given in the previous section.  In Ref.\ \citen{Dia}, the worldsheet
gauge theory of D-strings ending on the D3-branes is identified with
the $SU(2)$ monopole moduli space of ADHMN construction\cite{nahm} in
a 4 dimensional super-Yang-Mills-Higgs system. This identification
enables us to study a multi-pronged string whose component strings end
on D3-branes from the D-string worldsheet point of view.

\subsection{1/4 BPS state and energy bound}

To describe the perpendicular D3-branes in the D-string worldsheet
gauge theory, we turn on three scalar fields $X^i\equiv \xi^{i+2}$
($i=1,2,3$), which are interpreted as coordinates parameterizing the
D3-brane worldvolume. This means that the D3-branes extend in the
direction (0345). Together with the previously considered fields,
$A_0$ and $S$, the energy formula becomes
\begin{eqnarray}
\label{energy2}
\lefteqn{U
=\frac{T_{\rm D1}}{2}\int_{x^{(-)}}^{x^{(+)}}\!\! dx \;\tr
\Bigm[
  (\E\pm D_1S)^2+(D_0 S)^2\mp 2D_1S\E\qquad\qquad{}
}\nn\\
&&\qquad\qquad\qquad\qquad\qquad\!\!
+(D_0X^i)^2+(D_1X^i)^2 -[S,X^i]^2-\Half [X^i,X^j]^2
\!\Bigm]\!.
\end{eqnarray}
The gauge group is chosen to be $SU(2)$ for a reason that will be
explained below. The D-string terminates at $x=x^{(\pm)}$, where
the D3-branes are located. Thus this theory is defined on the spatial
interval $x^{(-)}<x<x^{(+)}$.

In the presence of the scalar fields $X^i$, the equation of motion for
the electric field reads
\begin{eqnarray}
\label{eqEX}
  D_1\E + i[X^i,D_0X^i]=J\delta(x-x_0),
\end{eqnarray}
where $J$ ($\equiv\sum_i j_i \sigma^i$) is a constant $SU(2)$ matrix
corresponding to the source $ng$ in Eq.\ (\ref{eqE}). With the help of
the resultant relation
\begin{eqnarray}
  \tr 2D_1S\E=\tr 2\p_1 (S\E) + i\tr 2[S,X^i]D_0X^i-2SJ\delta(x-x_0),
\end{eqnarray}
the energy (\ref{energy2}) is bounded in the same way, except for the
terms consisting only of the scalar fields $X^i$, as
\begin{eqnarray}
  U
&&=T_{\rm D1}\int\! dx \;\Half\tr
\Bigm[
  (\E\pm D_1S)^2+(D_0 S)^2\mp 2\p_1(S\E)\pm 2SJ\delta(x-x_0)
\nn\\
&&\qquad\qquad\qquad\quad
+\Bigl(D_0X^i\mp i[S,X^i]\Bigr)^2+(D_1X^i)^2 -\Half [X^i,X^j]^2
\Bigm].
\end{eqnarray}
For the last two terms consisting of $X^i$, we have the identity
\begin{eqnarray}
\lefteqn{
\tr\left[  (D_1X^i)^2 -\Half [X^i,X^j]^2\right]
}\nn\\
&&
\qquad
=\tr\sum_i
\left(
  D_1X^i\pm\frac{i}{2} \sum_{j,k}\epsilon^{ijk}X^jX^k
\right)^2
\mp \frac{2i}{3}\p_1\tr(\epsilon^{ijk}X^iX^jX^k).
\end{eqnarray}
Hence, finally the energy is bounded as follows:
\begin{eqnarray}
\label{enebou}
U
&&=T_{\rm D1}\int\! dx \;\Half\tr
\left[
  (\E\pm D_1S)^2+(D_0 S)^2+\Bigl(D_0X^i\mp i[S,X^i]\Bigr)^2 
\right.\nn\\
&&
\qquad\qquad\qquad\qquad\qquad\qquad
+\left.
\left(
  D_1X^i\pm\frac{i}{2}\epsilon^{ijk}X^jX^k
\right)^2\right]
\nn\\
&&\qquad\qquad
\mp T_{\rm D1}\tr\Bigl[ S\E\Bigr]_{x^{(-)}}^{x^{(+)}}
\pm 2T_{\rm D1}\tr\Bigl(JS(x\!=\!x_0)\Bigr)
\mp T_{\rm D1}\frac{i}{3}
\tr\Bigl[\epsilon^{ijk}X^iX^jX^k\Bigr]_{x^{(-)}}^{x^{(+)}}\nn\\
&&
\geq
\bigm| \E_{\rm boundary}+\E_{\rm source}\bigm|+\E_{\rm D3}
\end{eqnarray}
where
\begin{eqnarray}
\label{ED3}
  \E_{\rm D3}=T_{\rm D1}\frac{\mp i}{3}
\tr\Bigl[\epsilon^{ijk}X^iX^jX^k\Bigr]_{x^{(-)}}^{x^{(+)}}.
\end{eqnarray}

Therefore, the conditions for saturating the energy bound are, in
addition to (\ref{BPSS}), 
\begin{eqnarray}
  &&D_0X^i\mp i[S,X^i]=0, \label{susy}\\
  &&D_1X^i\pm\epsilon^{ijk}X^jX^k=0.\label{nahm}
\end{eqnarray}
In order to solve these BPS equations, (\ref{BPSS}), (\ref{susy}) and
(\ref{nahm}), first note that Eq.\ (\ref{susy}) is trivially
satisfied with the previous BPS condition (\ref{upper}), which is a
solution of the conditions (\ref{BPSS}). Thus we are left with 
(\ref{nahm}). This remaining condition is the Nahm
equation\cite{nahm} \ if we substitute the gauge fixing condition
$A_1=0$. 

In Ref.\ \citen{Dia}, the relation between this Nahm equation (included
in Nahm data) and the D-string approach in string theory was 
discussed. To precisely compare our situation with the Nahm data, we
choose the interval defining the worldsheet to satisfy
\begin{eqnarray}
  x^{(\pm)}=\pm\frac{\pi}{2a}
\end{eqnarray}
without loss of generality. D3-branes are located at these boundaries, 
$x=x^{(\pm)}$. It is claimed in Ref.\ \citen{Dia} (for related
discussion, see Ref.\ \citen{BC}) that the boundary conditions which
represent $k$ finitely separated D-strings terminating on the D3-branes
are
\begin{eqnarray}
\label{boucon}
  \wt{X}^i\sim\frac{T_i}{ \frac{\pi}{2}\mp z},\qquad 
   \left(z\sim\pm  \frac{\pi}{2} \right)
\end{eqnarray}
where new rescaled variables are defined as 
$\wt{X}^i \equiv \mp iX^i/a$, $z\equiv ax$, and the matrices $T^i$
define an irreducible $k$-dimensional representation of $SU(2)$:
\begin{eqnarray}
\label{T}
    [T^i,T^j]=\epsilon^{ijk}T^k.
\end{eqnarray}

The asymptotic expression (\ref{boucon}) with (\ref{T}) is actually a
solution of the Nahm equation (\ref{nahm}) near the boundaries. The 
scalar $X^i$ diverges at the boundaries, and this implies that there are
D3-branes there. Equation (\ref{boucon}) is consistent with the fact
that in the D3-brane worldvolume gauge theory,\cite{CM} an attached
D-string is represented as $|x-x^{(\pm)}|\sim 1/r$ with $r\sim |X^i|$,
at $r\sim\infty$.

It should be noted that for $k=1$ the solution of the Nahm equation is
trivially a constant.\footnote{With this Nahm data, one can construct
  a single BPS monopole solution\cite{BPS} using the ADHMN method.}
Extending this trivial solution, one finds that a constant diagonal
matrix also satisfies the Nahm equation (\ref{nahm}). Though this
vacuum is ordinarily adopted, it belongs to a reducible representation
of $SU(2)$ in the Nahm language. The author of Ref.\ \citen{Dia}
asserts that the reducible representation indicates that D-strings
exist infinitely far from each other, and that a configuration of
finitely separated D-strings should satisfy an irreducible boundary 
condition. Furthermore, with a constant diagonal matrix solution, it
is impossible to incorporate the effect of the D3-branes. Thus we
take the simplest choice $k=2$ in the following. Therefore we
set $T^i=\sigma_i/2i$.

Let us check the consistency in terms of the energy bound. The term
$\E_{\rm D3}$ in the energy bound (\ref{enebou}) is expected to 
involve the energy of the D3-branes at the boundaries. In fact, 
estimation of (\ref{ED3}) by substituting Eq.\ (\ref{boucon}) leads us
to the relation
\begin{eqnarray}
\label{assymp}
\E_{\rm D3}=
T_{\rm D1}\frac{1}{2(x^{(+)}-x)^3}
                     \biggm|_{x\rightarrow x^{(+)}}
+\;
T_{\rm D1}\frac{1}{2(x-x^{(-)})^3}
                     \biggm|_{x\rightarrow x^{(-)}}.
\end{eqnarray}
This expression is independent of the length of the interval (the
parameter $a$), and diverges correctly as $r^3$, which indicates the
volume of the D3-brane.\footnote{The tension of the D3-brane following
  from  Eq.\ (\ref{assymp}) differs from the usual tension $T_{\rm
    D3}(=T_{\rm D1})$, by a factor 2. This unsatisfactory
  disagreement may originate from the fact that we must treat at least
  two D-strings to represent the D3-branes as boundary conditions, as
  explained above.}

The Nahm equation (\ref{nahm}) with non-trivial boundary
conditions breaks half of the original supersymmetry.\cite{Dia} \
Since we have other BPS equations, (\ref{BPSS}) and (\ref{susy}), in
addition to (\ref{nahm}), the preserved supersymmetry expected in our
case is 1/4 of the original one. Using the gaugino transformation
\begin{eqnarray}
   \delta\lambda_{\alpha i}=&&2
  \left[
    (\sigma^0\bar{\sigma}^1-\sigma^1\bar{\sigma}^0)_\alpha^{\;\beta}
    \zeta_{\beta i} \E
    +
    (\sigma^0\bar{\sigma}^3-\sigma^3\bar{\sigma}^0)_\alpha^{\;\beta}
    \zeta_{\beta i} D_0S
    +
    (\sigma^1\bar{\sigma}^3-\sigma^3\bar{\sigma}^1)_\alpha^{\;\beta}
    \zeta_{\beta i} D_1S
  \right]\nn\\
&&-4i\sigma^\mu_{\alpha\dot{\alpha}}(\zeta_{\beta j})^*
\epsilon^{\dot{\beta}\dot{\alpha}}(\wt{\sigma}_m)_{ij}D_\mu X^m
-4\sigma^3_{\alpha\dot{\alpha}}(\zeta_{\beta j})^\dagger
\epsilon^{\dot{\beta}\dot{\alpha}}(\wt{\sigma}_m)_{ij}[S,X^m]
\nn\\
&&+2i(\wt{\sigma}_m\wt{\sigma}^*_n)_{ij}\zeta_{\alpha j}[X^m,X^n],
\end{eqnarray}
we find that the actual supersymmetry surviving under all the BPS
conditions, (\ref{BPSS}), (\ref{susy}) and (\ref{nahm}), is 1/4 of the
original one. The unbroken supersymmetry is generated by a
supersymmetric transformation parameter $\zeta_j$ satisfying
(\ref{1/2}) and
\begin{eqnarray}
  \zeta_j = \mp i \sigma^1_{\alpha\dot{\alpha}}
            \epsilon^{\dot{\beta}\dot{\alpha}}(\zeta_{j\beta})^*
            M_{ij}
\qquad\mbox{where}\quad
M=
\left(\begin{array}{cccc}
 & & &-1\\
 & &1&  \\
 &1& &  \\
-1& & & \\              
\end{array}\right).
\end{eqnarray}

\subsection{Shape of the three-pronged string}
\setcounter{footnote}{0}

In this subsection, we consider configurations of the
three-pronged string with the presence of D3-branes. The shape of
the pronged string is determined by the scalar field $S(x)$. Since the
variables $X^i$ are already mutually non-commutative, direct 
interpretation of the locations of the D-strings is rather ambiguous,
as discussed in Ref.\ \citen{Dia}. Here, for the scalar $S$ we turn on
only its diagonal part, and observe how it behaves with the
presence of D3-branes.

Due to one of the BPS conditions ($A_0\!=\!S$), the scalar field $S$
obeys the Gauss law constraint (\ref{eqEX}):
\begin{eqnarray}
\label{solve}
    \p_1^2 S + i\sum_i[X^i,[S,X^i]]=-J\delta(x-x_0).
\end{eqnarray}
For the background configuration $X^i$ of the D3-branes, we adopt a
solution of the Nahm equation which was given in Ref.\ \citen{mono}.
The authors of Ref.\ \citen{mono} explicitly solved the Nahm equation
for $k=2$, obtaining
\begin{eqnarray}
\label{twmo}
  \wt{X}^i\equiv f_i(z)\frac{\sigma_i}{2i} \qquad \mbox{for} \;\;
  i=1,2,3,
\end{eqnarray}
where $\sigma_i$ are Pauli matrices and 
\begin{eqnarray}
\label{resca}
f_1(z)=A\frac{{\sn} (u,k)}{{\cn} (u,k)}, \quad
f_2(z)=A\frac{1}{{\cn} (u,k)},\quad
f_3(z)=A\frac{1}{k'}\frac{{\dn} (u,k)}{{\cn} (u,k)},
\end{eqnarray}
with $u\equiv\frac{2Kz}{\pi}$. The constant numbers $A, k', K$ and $k$
are defined as 
\begin{eqnarray}
&&A(k)\equiv\frac{2k' K}{\pi}, \quad 
k'(k)\equiv\sqrt{1-k^2},\nn\\
&&K(k)\equiv\int^{\pi/2}_0\frac{dy}{\sqrt{1-k^2\sin^2 y}},\quad 
k\equiv\frac{\delta}{\sqrt{1+\delta^2}}.
\end{eqnarray}
In the ADHMN construction of monopoles, this solution\footnote{The
  solution (\ref{resca}) represents a rescaling of the solution given
  in Ref.\ \citen{mono}, so that the solution has poles at $z=\pm
  \frac{\pi}{2}$.} 
represents two finitely separated monopoles that are aligned along the
$X^3$ axis. The single parameter $\delta$ in the solution denotes
the distance between the two monopoles. For $\delta=0$, two monopoles
are located at the same point and we regain the axially symmetric 
two-monopole solution given in Ref.\ \citen{ax}. On the other hand, for a
large $\delta$ the distance between the two diverges as $\log \delta$.
Thus it is expected that the large $\delta$ limit leads us to the
situation in which the two D-strings are decoupled from each other.

Let us solve the second order differential equation (\ref{solve}).
Decomposing $S(x)$ into its adjoint components $s_i(x)$ as 
$S(x)\equiv\sum_{i}s_i(x)\sigma^i$, and substituting this and Eq.\
(\ref{twmo}) into (\ref{solve}), we obtain a differential equation for
$x\neq x_0$:
\begin{eqnarray}
\label{fj2}
  \left(
\frac{d^2}{dx^2}- a^2\sum_{j(\neq i)}f_j^2
  \right)
s_i=0.
\end{eqnarray}
Since the boundary behavior of the potential term is already known, as
in Eq.\ (\ref{boucon}), one can solve the differential equation
(\ref{fj2}) near the boundary $x\sim x^{(+)}$ (or $x^{(-)}$). The only
regular solution is $s \sim (x-x^{(+)})^2$. (The other solution
satisfies found to be divergent as $s \sim (x-x^{(+)})^{-1}$.) Without
use of the source term $J$, it is impossible to have a solution
regular at both boundaries $x^{(+)}$ and $x^{(-)}$, since the
potential term $f_j^2$ is positive definite.

As mentioned above, we consider only the differential equation for
$i=3$,
\begin{eqnarray}
\label{diff1}
\left[
  \left(
    \frac{d}{dz}
  \right)^2 -(f_1^2 +f_2^2)
\right]s_3(z)=0,
\end{eqnarray}
and according to this, the components of the source $J$ are chosen as 
\begin{eqnarray}
  j_1=j_2=0,\qquad j_3=ng.
\end{eqnarray}

\begin{figure}[htdp]
\begin{center}
\begin{minipage}{65mm}
\begin{center}
\leavevmode
\epsfxsize=65mm
\epsfbox{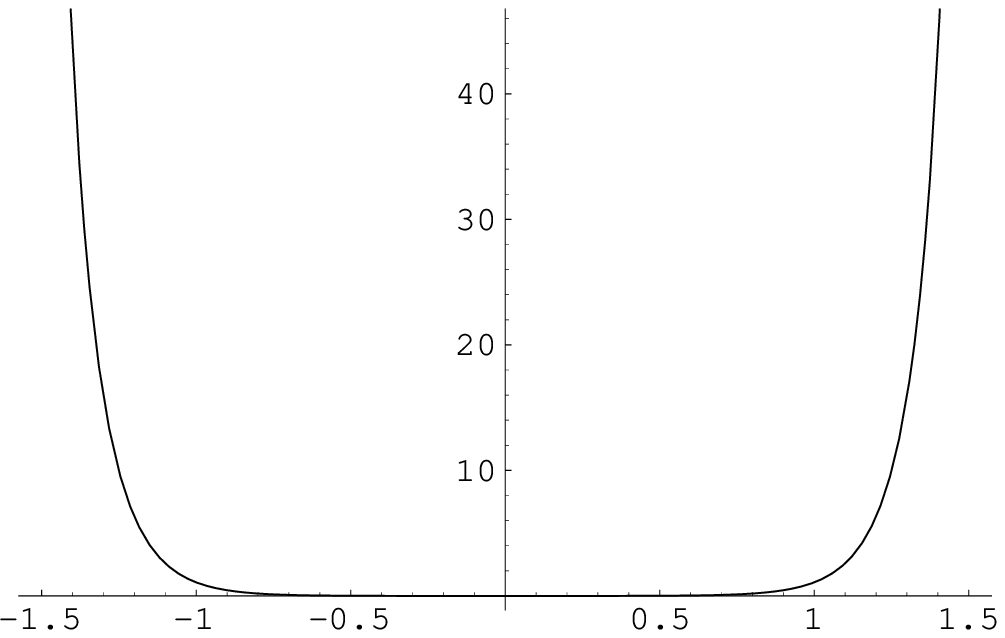}
\put(-15,15){$z$}
\put(-120,125){$f_1^2(z) + f_2^2(z)$}
\caption{
The potential term $f_1^2(z) + f_2^2(z)$ with large $\delta$ ($\sim
700$), which vanishes except near the boundaries.}
\label{fig:po}
\end{center}
\end{minipage}
\hspace{5mm}
\begin{minipage}{65mm}
\begin{center}
\leavevmode
\epsfxsize=65mm
\epsfbox{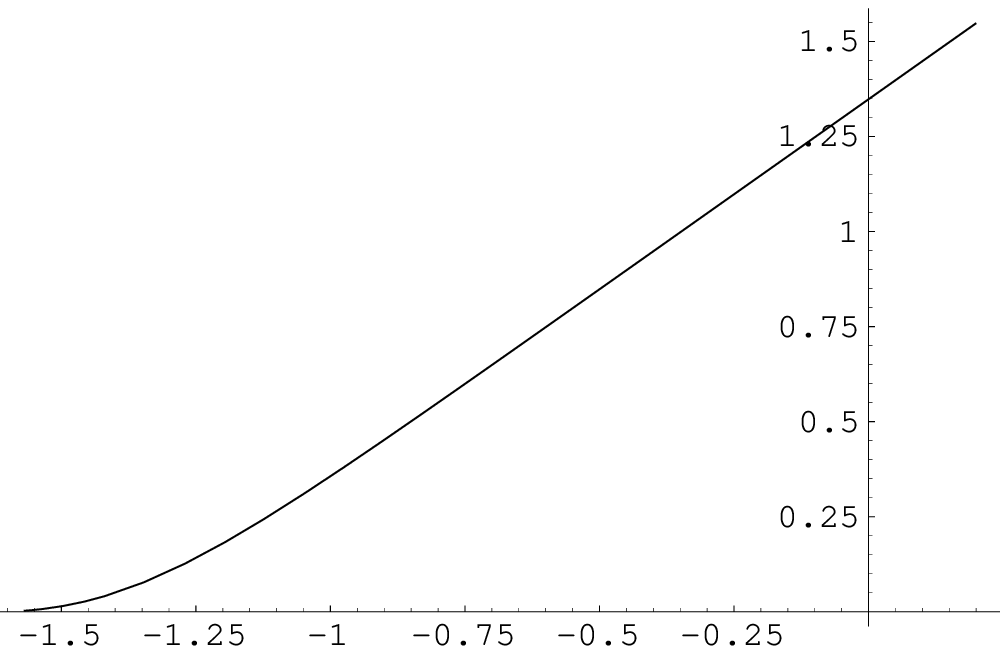}
\put(-15,15){$z$}
\put(-35,130){$s_3(z)$}
\caption{A solution of Eq.\ (\ref{diff1}) regular at one boundary
  $z=-\pi/2$ for a large $\delta$. One can observe that the solution
  is almost a straight line.}  
\label{fig:so}
\end{center}
\end{minipage}
\end{center}
\end{figure}

First, consider the large separation (large $\delta$) case. We
estimate the potential term in Eq.\ (\ref{diff1}) as follows. For a
given positive $z (\neq \pi/2)$, the solution $f_1,f_2$ presented in
Eq.\ (\ref{resca}) satisfies
\begin{eqnarray}
  f_1^2 + f_2^2 \sim \frac{1}{\pi^2}(\log \delta)^2
  \delta^{-2(1-2z/\pi)}
\longrightarrow 0 \qquad \mbox{for}\quad \delta\longrightarrow \infty.
\end{eqnarray}
This is a natural result: when we separate two monopoles (D-strings),
the interaction between the two vanishes and the situation would
reduce to the two decoupled D-string ($U(1)\times U(1)$) case. 
Actually, for $\delta\sim 700$, a numerical estimation for the
potential (see Fig.\ \ref{fig:po}) shows that the potential vanishes
except near the boundaries. Therefore, in the large $\delta$ limit,
the solution of the differential equation (\ref{diff1}) with the
background (\ref{resca}) is found to be almost of constant slope, as
shown in Fig.\ \ref{fig:so}. Thus we conclude that the large $\delta$
limit recovers the picture with no D3-brane discussed in section
\ref{2}, where 
the component strings are straight.

The point here which differs from the situation in section \ref{2} is
that, at the 
boundaries $z=\pm \pi/2$, the end of string 1 and 3 are constrained as 
$s_3(z\!=\!\pm \pi/2)=0$. In the infinite $\delta$ limit, the potential
vanishes except at the boundaries and the solution of the differential
equation becomes
\begin{eqnarray}
\label{combi}
s_3(z)=
 \left\{
  \begin{array}{ll}
    C^{(-)}s^{(-)}(z), \qquad&(z<z_0)\\
    C^{(+)}s^{(+)}(z), &(z>z_0)
  \end{array}
\right.
\end{eqnarray}
where $z_0\equiv ax_0$ and the component strings are straight:
\begin{eqnarray}
  s^{(+)}(z)=z-\frac{\pi}{2},\qquad s^{(-)}(z)=z+\frac{\pi}{2}.
\end{eqnarray}
Continuity of the solution at $z=z_0$ requires
\begin{eqnarray}
\label{continu}
  \frac{C^{(+)}}{C^{(-)}}=\frac{z_0+\frac{\pi}{2}}{z_0-\frac{\pi}{2}}.
\end{eqnarray}
Information regarding the source $J$ is incorporated in the boundary 
condition at $x=x_0$, concerning the derivative of $S$. Actually, this
refrects the charge conservation condition at the junction point
$x=x_0$ as seen in section \ref{2}. If one regards the source charge
$n$ as that which the invisible fundamental string (denoted as
string 2) possesses, and noting that the charge of each string 1 and 3
is given by the slope of the string measured at $x=x_0$,
$-\frac{d}{dx}S\bigm|_{x=x_0}$, the charge conservation condition at
the junction point implies
\begin{eqnarray}
\label{cc}
 -pg = -C^{(-)}\frac{d}{dx}s^{(-)},\qquad
ng-pg = -C^{(+)}\frac{d}{dx}s^{(+)}.
\end{eqnarray}
Given a pair of charges $p$ and $n$, the boundary conditions
(\ref{continu}) and (\ref{cc}) determine the solution perfectly.  

Second, we consider the case of two monopoles with coincident
locations, {\it i.e.}, $\delta=0$. The elliptic functions appearing in 
(\ref{resca}) are simplified to usual trigonometric functions as
\begin{eqnarray}
  f_1=f_3=\frac1{\cos z},\qquad f_2=\tan z.
\end{eqnarray}
Then the differential equation (\ref{diff1}) becomes
\begin{eqnarray}
\label{diff}
\left[
  \left(
    \frac{d}{dz}
  \right)^2 -\left(\frac{2}{\cos^2z}-1\right)
\right]s_3(z)=0.
\end{eqnarray}
Changing the variable as $w\equiv \cos^2 z$ and defining $\wt{s}\equiv 
s/w$, this differential equation is then transformed into a Gauss
\begin{figure}[tdp]
\begin{center}
\begin{minipage}{60mm}
\begin{center}
\leavevmode
\epsfxsize=60mm
\epsfbox{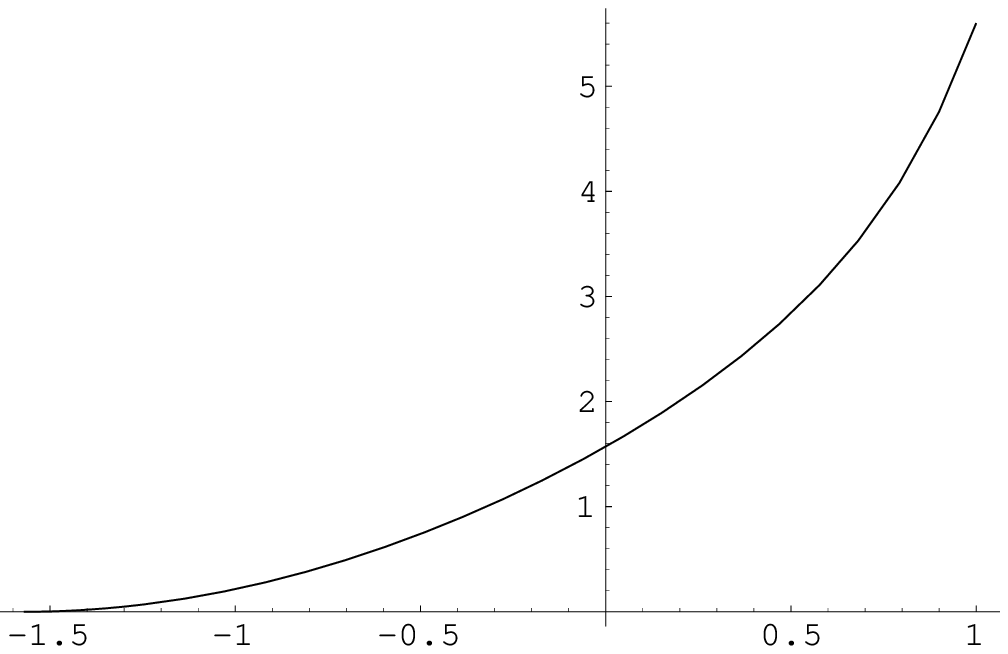}
\put(-10,15){$z$}
\put(-85,120){$s^{(-)}(z)$}
\caption{A solution $s^{(-)}(z)$, which is regular at $z=-\pi/2$.}
\label{fig:s-}
\end{center}
\end{minipage}
\hspace{5mm}
\begin{minipage}{60mm}
\begin{center}
\leavevmode
\epsfxsize=60mm
\epsfbox{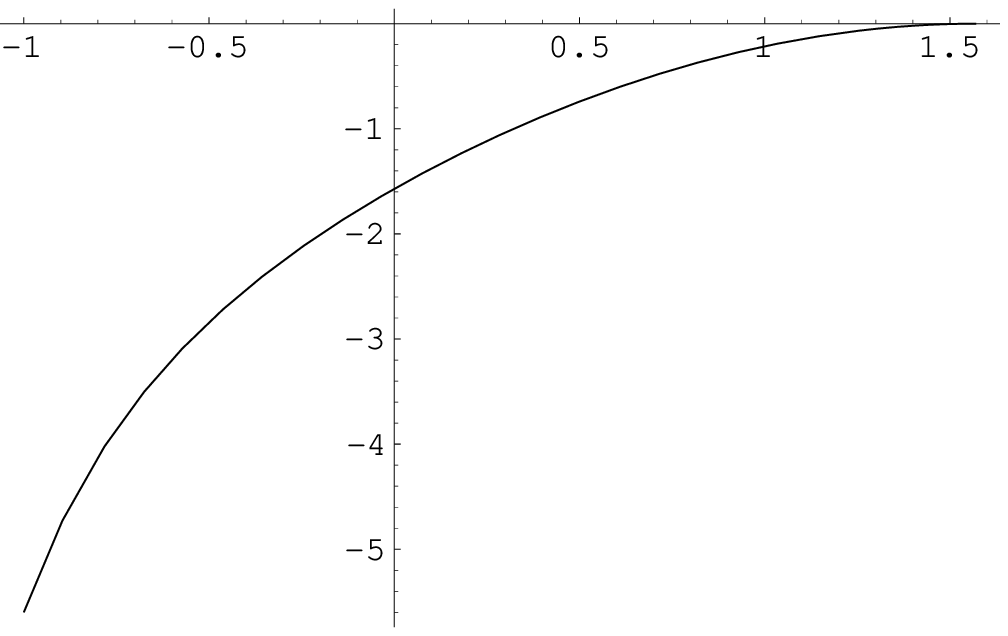}
\put(-160,110){$z$}
\put(-90,10){$s^{(+)}(z)$}
\caption{A solution $s^{(+)}(z)$, which is regular at $z=\pi/2$.}
\label{fig:s+}
\end{center}
\end{minipage}
\end{center}
\end{figure}
hypergeometric differential equation. Using some formulas concerning
hypergeometric functions, we obtain two independent solutions of
(\ref{diff}),
\begin{eqnarray}
  s^{(\pm)}(z)=\sin z+
  \left(
    z\mp\frac{\pi}{2}
  \right)\frac1{\cos z}.
\end{eqnarray}
This solution $s^{(+)}(z)$ ($s^{(-)}(z)$) is regular at the boundary
$z=\pi/2$ ($z=-\pi/2$), whose shape is plotted in  Fig.\ \ref{fig:s-}
(Fig.\ \ref{fig:s+}).
Hence we get a configuration that is regular in the entire region, as
for Eq.\ (\ref{combi}). In this case, the continuity of the
solution at $z=z_0$ requires
\begin{eqnarray}
\label{cont}
  \frac{C^{(+)}}{C^{(-)}}=\frac{\sin z_0+
    \left(
      z_0+\frac{\pi}{2}
    \right)\frac1{\cos z_0}}
   {\sin z_0+
    \left(
      z_0-\frac{\pi}{2}
    \right)\frac1{\cos z_0}}.
\end{eqnarray}
Solving the charge conservation condition (\ref{cc}) at $z=z_0$
together with Eq.\ (\ref{cont}), we have a complete form of the
pronged string.

\begin{figure}[tdp]
\begin{center}
\begin{minipage}{6cm}
\begin{center}
\leavevmode
\epsfxsize=60mm
\epsfbox{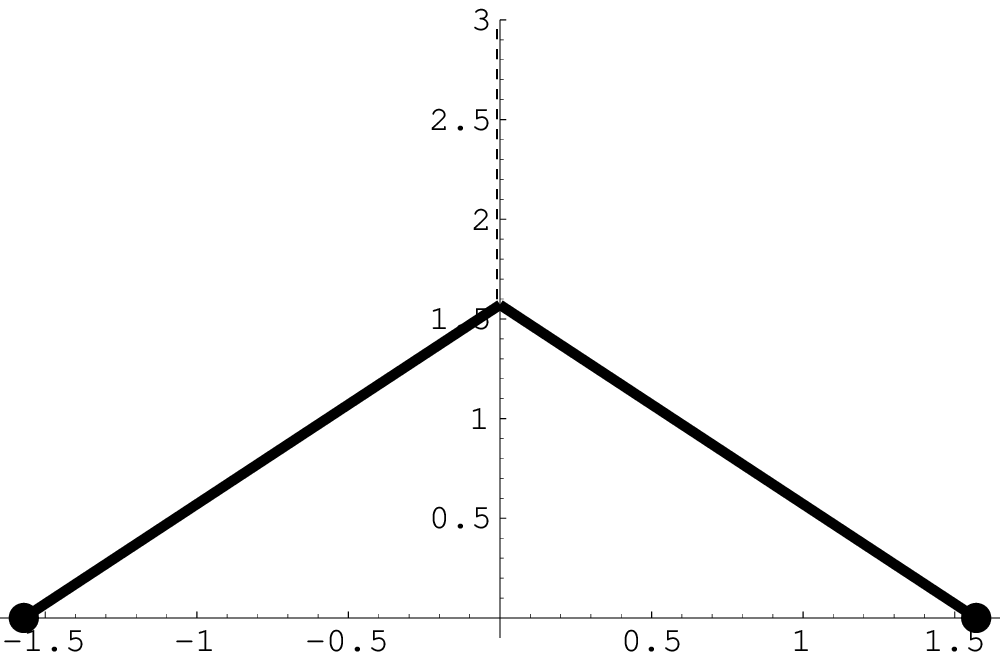}
\put(0,20){$z$}
\put(-90,120){$s(z)$}
\caption{The case $n=2$ and $p=1$, for $\delta=\infty$. This
  corresponds to the [$(-1,1)$, $(2,0)$, $(-1,-1)$] junction. The
  dashed line (on the line $z=0$) denotes the string 2. The two blobs
  indicate the locations of the D3-branes.}  
\label{fig:bend3}
\end{center}
\end{minipage}
\hspace*{5mm}
\begin{minipage}{6cm}
\begin{center}
\leavevmode
\epsfxsize=60mm
\epsfbox{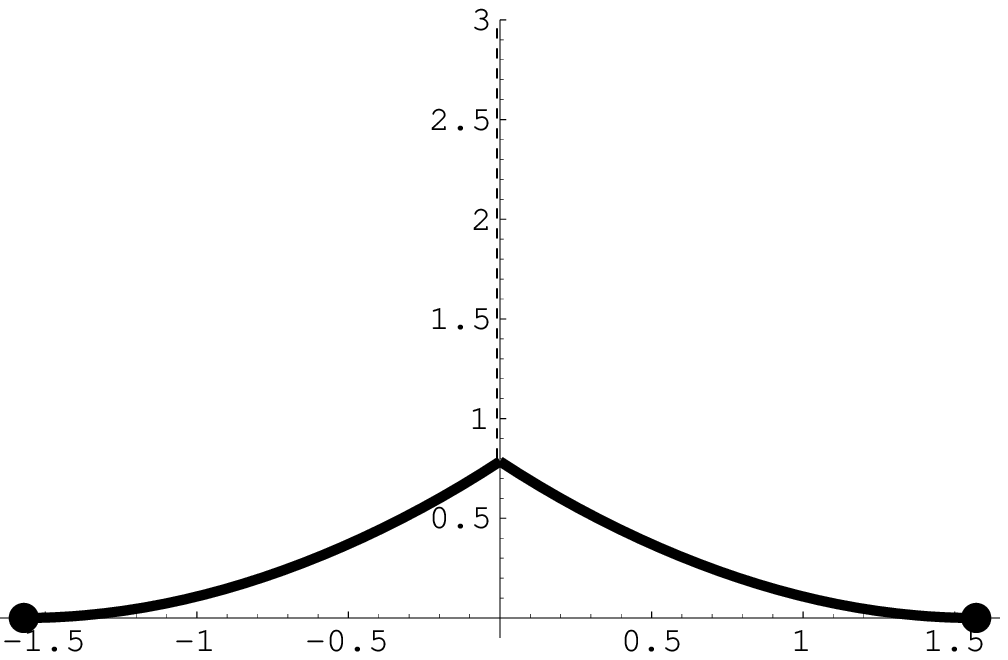}
\put(0,20){$z$}
\put(-90,120){$s(z)$}
\caption{The case $n=2$ and $p=1$, with $\delta=0$. This
  corresponds to the [$(-1,1)$, $(2,0)$, $(-1,-1)$] junction. The
  string 2 is also on the line $z=0$.}
\label{fig:bend1}
\end{center}
\end{minipage}
\end{center}
\end{figure}

\begin{figure}[tdp]
\begin{center}
\begin{minipage}{6cm}
\begin{center}
\leavevmode
\epsfxsize=60mm
\epsfbox{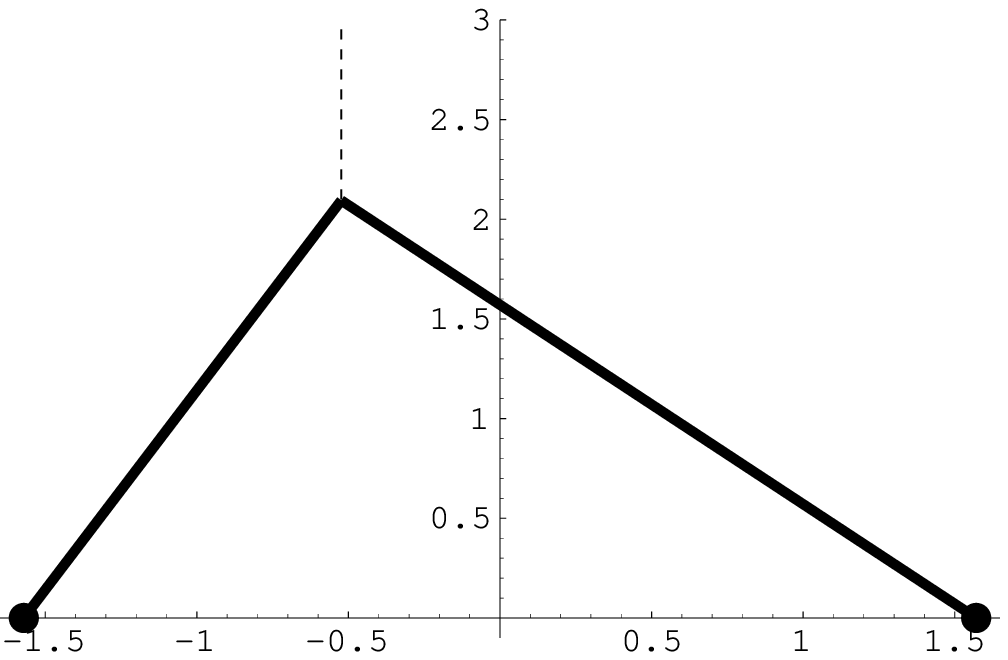}
\put(0,20){$z$}
\put(-90,120){$s(z)$}
\caption{For $n=3$ and $p=2$ ([$(-2,1)$, $(3,0)$, $(-1,-1)$]
  junction) with $\delta=\infty$, the junction point is not on the
  axis $z=0$. } 
\label{fig:bend4}
\end{center}
\end{minipage}
\hspace*{5mm}
\begin{minipage}{6cm}
\begin{center}
\leavevmode
\epsfxsize=60mm
\epsfbox{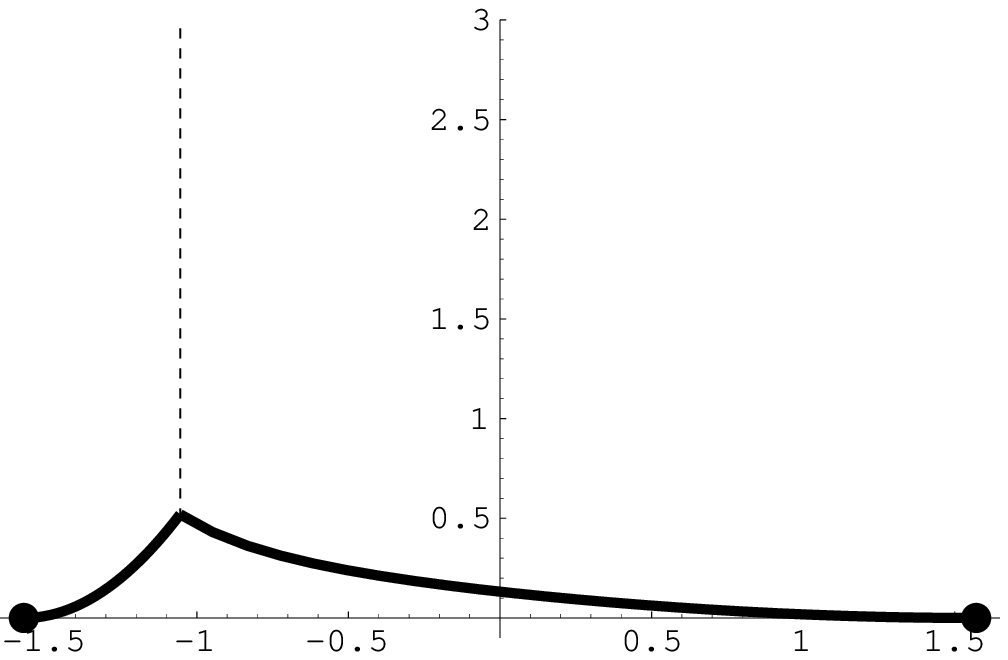}
\put(0,20){$z$}
\put(-90,120){$s(z)$}
\caption{For $n=3$ and $p=2$ ([$(-2,1)$, $(3,0)$, $(-1,-1)$]
  junction), with $\delta=0$.} 
\label{fig:bend2}
\end{center}
\end{minipage}
\end{center}
\end{figure}

Here we explicitly analyze two typical cases.
\begin{itemize}
\item Case (1) : 
[$(-1,1)$, $(2,0)$, $(-1,-1)$] junction ($n=2$ and $p=1$). Parity
invariance of  the boundary conditions indicates that the junction
point sits at $x=0$, and the solution of the boundary conditions at
the junction point (the continuity condition (\ref{cont}) and the
charge conservation (\ref{cc})) is found as follows:
\begin{eqnarray}
  &\mbox{For} \;\; \delta=\infty,
   \qquad &C^{(+)} = - C^{(-)} = -\frac{g}{a}.\nn\\
  &\mbox{For} \;\; \delta=0,
   \qquad &C^{(+)} = - C^{(-)} = -\frac{g}{2a}.
\end{eqnarray}
This configuration is plotted in Fig.\ \ref{fig:bend1}
($\delta=\infty$) and Fig. \ref{fig:bend3} ($\delta=0$). 
\item Case (2) : 
[$(-2,1)$, $(3,0)$, $(-1,-1)$] junction ($n=3$ and $p=2$). In this
case there is no parity invariance:
\begin{eqnarray}
  \mbox{For} \;\; \delta=\infty,
   \quad & C^{(+)} = -\displaystyle \frac{g}{a},\;\;C^{(-)} =
   \frac{2g}{a},\;\; 
   z_0=-\frac{\pi}{6}.
\end{eqnarray}
For the case $\delta=0$, a numerical computation gives the 
solution of Eqs.\ (\ref{cc}) and (\ref{cont}) as 
\begin{eqnarray}
C^{(+)} = -0.084\,\frac{g}{a},\;\; C^{(-)} = 2.957\,\frac{g}{a},
\;\;z_0=-0.167.
\end{eqnarray}
The shape of the pronged-string is shown in Fig.\ \ref{fig:bend2} 
($\delta=\infty$) and Fig. \ref{fig:bend4} ($\delta=0$).
\end{itemize}
In both cases, there must be invisible fundamental strings stretched
from the junction point upward. They are depicted with dashed lines
in the figures. 

For the infinite $\delta$ case we observe that the component strings
are straight due to the vanishing of potential terms representing the
D3-branes. On the other hand, when the locations of the two monopoles 
coincide ($\delta=0$), one can see in Figs.\ \ref{fig:bend1} and
\ref{fig:bend2} that the D-string trajectories bend non-trivially. 
The interpretation is as follows: 
since we have considered the $\sigma_3$
component of the matrix $S(x)$, there is also another D-string
(pronged string) given by $-s_3(x)$ which has charges of the
opposite 
\begin{wrapfigure}{r}{70mm}
\begin{center}
\leavevmode
\epsfxsize=70mm
\epsfbox{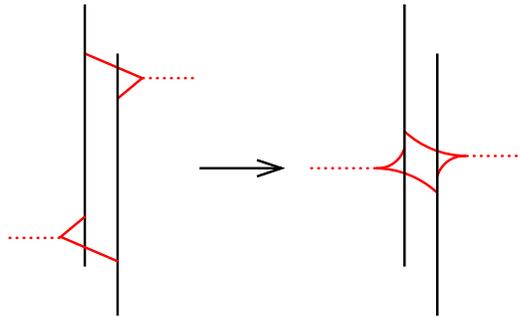}
\caption{Two three-pronged strings interacting with each other through
  the common D3-branes.}
\label{fig:two}
\end{center}
\end{wrapfigure}
sign, in addition to the pronged string represented by
$s_3(x)$. The bending of the strings seems to be due to the 
interaction (attractive force) between these two pronged strings. In
other words, from the point of view of D-string worldsheet gauge
theory, the two diagonal elements of $S(x)$ interact with each other
through the non-diagonal elements of the other scalar fields
$X^i(x)$. This implies that the interaction between the two string 
junctions is through the D3-branes at the boundaries. A sketch of this
configuration is presented in Fig.\ \ref{fig:two}.

\section{Summary and discussion}
\label{4}

In this paper, we have studied the 1/4 BPS states in the D-string
worldsheet gauge theory. These states correspond to the multi-pronged
strings in the string picture. We have combined the ideas presented in
Ref.\ \citen{DASG} (where the infinitely long pronged string is
realized in the worldsheet gauge theory) and Ref.\ \citen{Dia} (in
which the D-strings terminating on the D3-branes are considered as a
notable realization of the Nahm equation). This enables us to treat a
pronged string ending on the D3-branes, from the D-string worldsheet
point of view.

First, in section \ref{2}, we studied the energy of the
infinitely long three-pronged strings. The saturated BPS energy bound 
ensures the interpretation of the kink solution as a pronged string. 
In particular, there is actually an invisible fundamental string
extending from the junction point.

Second, in section \ref{3}, we deduced BPS conditions which
preserve 1/4 supersymmetry. One of the BPS equations is the Nahm equation, 
and the Nahm boundary condition is found to be consistent with the
investigation of the energy at boundaries where D3-branes are
located. Then we solved the BPS equations explicitly for a background
configuration of two monopoles in two limits: (i) the two monopoles 
are infinitely separated, and (ii) locations of the two monopoles
coincident. For the case (i), in which physically we should regain a
configuration of two decoupled pronged strings, we have found that
component strings forming the junction are straight. Hence, as
expected, the configurations agree with the naive prediction in which
the situation would be reduced to the case of no D3-brane. On the
other hand, for the case (ii), the resultant shape of the
three-pronged strings is found to be nontrivial in the sense that
component strings bend. This seems to reflect the fact that we have
inevitably considered two D-strings ($SU(2)$ gauge group), and these
two interact with each other through the D3-branes.

The obtained shape of the three-pronged string is not similar to the
one given in Ref.\ \citen{ours2}, where at the junction point two of
the three strings which possess magnetic charges connect smoothly. 
Furthermore, for our configuration studied in section \ref{3} with
$\delta\neq\infty$, the tensions of three strings are balanced only at
the junction point, in contrast to the configuration in Ref.\
\citen{ours2}. This discrepancy should be examined in more detail
for two reasons. One is a generalization of the Nahm construction: The
ADHMN construction of monopoles has a splendid 
realization, a worldsheet theory of D-strings ending on D3-branes. Now
that we know the string junction configuration in the string picture,
what is the corresponding way of constructing the 1/4 BPS
states in $d\!=\!4$ supersymmetric Yang-Mills theory recently
presented in Refs.\ \citen{ours} -- \citen{LY}?
The other reason is of purely geometrical interest. The D-string
approach may provide information regarding a method we can use to see
the singular junction point.

There are subtle points when we conclude that strings bend. As
discussed in Ref.\ \citen{Dia}, the background configuration of
D3-branes is represented by the variables $X^i$ which do not commute
with each other. Therefore the coordinate 
interpretation is problematic. It is uncertain whether the bending
solutions given in section 3 can be regarded as a ``real''
configuration of the three-pronged strings.
Another problem is that we have dealt with marginal bound states of
fundamental strings named `string 2'. Unfortunately, in our scheme it
is difficult to treat a single fundamental string as string 2, since
in that case we have only a divergent solution. How to incorporate
general brane configurations is left as a future problem.


\section*{Acknowledgements}

I would like to thank N.\ Sasakura, B.\ -H.\ Lee and S.\ Imai
for helpful comments, and gratefully thank H.\ Hata for
valuable discussions.

\appendix
\section{Non-Abelian Born-Infeld Soliton}

At low energy, D-brane dynamics is precisely determined by the D-brane
action, whose gauge group usually taken into account is Abelian. 
However, intrinsically the gauge group necessary for our treatment in
this paper is non-Abelian. Thus if we wish to investigate precisely
the dynamics of D-strings we should make use of the non-Abelian
Born-Infeld (BI) action proposed in Ref.\ \citen{Tseytlin}. Here we
show that the BPS solutions satisfying the BPS equations
(\ref{BPSS}), (\ref{susy}) and (\ref{nahm}) are also solutions of the
non-Abelian BI equations of motion.

This subject was studied for the first time in Ref.\
\citen{Hashi} concerning the monopole configuration in $SU(2)$ super
Yang-Mills theory. In Ref.\ \citen{Bre}, various 1/2 BPS objects
satisfying non-Abelian BI equations of motion are collected. For the
1/4 BPS states, the authors of Ref.\ \citen{ours2} have proven this
property, using the techniques developed in Ref.\ \citen{Hashi}. The
method for verification adopted in this appendix is exactly the same
as that in Refs.\ \citen{ours2} and \citen{Hashi}.

We have four scalars, $S$ and $X^i$ Thus the low energy effective
action is deduced simply by dimensional reduction from the
$6$-dimensional BI action\footnote{We have ignored an overall
  numerical factor, which is irrelevant here, and have appropriately
  normalized the gauge fields to be consistent with Eq.\
  (\ref{action}).}
with a flat metric:\cite{Gibb}
\begin{eqnarray}
  S = \int\! d^6x\;\sqrt{-\det (\eta_{ab}+F_{ab})}.
\label{act6}
\end{eqnarray}
Here the fields representing the fluctuations of the D-string are
identified with some components of the six-dimensional gauge fields in
(\ref{act6}) through T-duality as $S=A_2$, $X^i=A_{i+2}$ ($i=1,2,3$). 
This identification results in the $6$-dimensional field strength as
follows: 
\begin{eqnarray}
&&F_{ab}=\nn\\[10pt]
&&\hspace*{-5pt}\pmatrix{
0      &\!\! \E     &\!\! D_0S     &\!\! D_0X^1     
&\!\! D_0X^2     &\!\! D_0X^3    \cr
-\E    &\!\! 0      &\!\! D_1S     &\!\! D_1X^1     
&\!\! D_1X^2     &\!\! D_1X^3    \cr
-D_0S  &\!\! -D_1S  &\!\! 0        &\!\! i[S,X^1]   
&\!\! i[S,X^2]   &\!\! i[S,X^3]  \cr
-D_0X^1&\!\! -D_1X^1&\!\! -i[S,X^1]&\!\! 0          
&\!\! i[X^1,X^2] &\!\! i[X^1,X^3]\cr
-D_0X^2&\!\! -D_1X^2&\!\! -i[S,X^2]&\!\! -i[X^1,X^2]
&\!\! 0          &\!\! i[X^2,X^3]\cr
-D_0X^3&\!\! -D_1X^3&\!\! -i[S,X^3]&\!\! -i[X^1,X^3]
&\!\! -i[X^2,X^3]&\!\! 0 }.
\label{fiestr}
\end{eqnarray}

A non-Abelian version of the BI action suffers from the ordering
ambiguities of the non-commutative field components. This ambiguity is
fixed in the proposal of Ref. \citen{Tseytlin} by taking the
symmetrized trace operation STr. The non-Abelian BI action is defined
by
\begin{eqnarray}
\label{nbi}
    S = \int\! d^6x\; {\rm STr} \sqrt{-\det (\eta_{ab}+F_{ab})}.
\end{eqnarray}

Using the explicit expression (\ref{fiestr}) and the BPS equations
(\ref{BPSS}), (\ref{susy}) and (\ref{nahm}), it is possible to reduce
the equations of motion from the action (\ref{nbi}) to the ordinary
Yang-Mills equations of motion,
\begin{eqnarray}
\label{eq:YMeq}
  D^aF_{ab}=0,
\end{eqnarray}
as in Refs.\ \citen{ours2} and \citen{Hashi}. Since the BPS
equations (\ref{BPSS}), (\ref{susy}) and (\ref{nahm}) satisfy the
Yang-Mills equations of motion (\ref{eq:YMeq}), we have shown that our
BPS saturated configuration is also a solution of the equations of
motion from the non-Abelian BI action. 

\newcommand{\J}[4]{{#1} {\bf #2} (#3), #4}
\newcommand{\andJ}[3]{{\bf #1} (#2) #3}
\newcommand{\AP}{Ann.\ Phys.\ (N.Y.)}
\newcommand{\MPL}{Mod.\ Phys.\ Lett.}
\renewcommand{\NP}{Nucl.\ Phys.}
\renewcommand{\PL}{Phys.\ Lett.}
\renewcommand{\PR}{Phys.\ Rev.}
\renewcommand{\PRL}{Phys.\ Rev.\ Lett.}
\renewcommand{\PTP}{Prog.\ Theor.\ Phys.}
\newcommand{\hepth}[1]{{\tt hep-th/#1}}


\begin{thebibliography}{99}
\bibitem{ours}
    K.\ Hashimoto, H.\ Hata and N.\ Sasakura,
      \J{\PL}{B431}{1998}{303}, \hepth{9803127}.

\bibitem{ours2}
    K.\ Hashimoto, H.\ Hata and N.\ Sasakura,
      \J{\NP}{B535}{1998}{83}, \hepth{9804164}.

\bibitem{kawano}
    T.\ Kawano and K.\ Okuyama, \J{\PL}{B432}{1998}{338},
    \hepth{9804139}.

\bibitem{LY}
    K.\ Lee and P.\ Yi,  \J{\PR}{D58}{1998}{066005}, \hepth{9804174}.

\bibitem{SHW}
    J.\ H.\ Schwarz,  \J{Nucl.\ Phys.\ Proc.\
      Suppl.}{55B}{1997}{1}, \hepth{9607201} ; \\
    O.\ Aharony, J.\ Sonnenschein and S.\ Yankielowicz,
    \J{\NP}{B474}{1996}{309}, \hepth{9603009}.

\bibitem{DASG}
    K.\ Dasgupta and S.\ Mukhi, 
    \J{\PL}{B423}{1998}{261},  \hepth{9711094}.

\bibitem{ZW}
    M.\ R.\ Gaberdiel and B.\ Zwiebach, 
    \J{\NP}{B518}{1998}{151}, {\tt hep-th/} {\tt 9709013} ; \\
    M.\ R.\ Gaberdiel, T.\ Hauer and B.\ Zwiebach, 
    \J{\NP}{B525}{1998}{117}, {\tt hep-th/} {\tt 9801205} ; \\
    Y.\ Imamura, 
    \J{\PR}{D58}{1998}{106005},  \hepth{9802189} ; \\
    O.\ DeWolfe and B.\ Zwiebach, \J{\NP}{B541}{1999}{509},
    \hepth{9804210}.

\bibitem{sen}
    A.\ Sen,  \J{JHEP}{9803}{1998}{005},  {\tt hep-th/9711130} ;  \\
    S.\ -J.\ Rey and J.\ -T.\ Yee, \J{\NP}{B526}{1998}{229} , {\tt
    hep-th/9711202}. 
    
\bibitem{CT}
    C.\ G.\ Callan and L.\ Thorlacius,
    {\tt hep-th/} {\tt 9803097}.

\bibitem{Gau}
    J.\ P.\ Gauntlett, J.\ Gomis and P.\ K.\ Townsend,
    \J{J.\ High Energy Phys.}{01}{1998}{003},  \hepth{9711205}.

\bibitem{KRO}
    M.\ Krogh and S.\ Lee, 
    \J{\NP}{B516}{1998}{241}, \hepth{9712050} ; \\ 
    Y.\ Matsuo and K.\ Okuyama, \J{\PL}{B426}{1998}{294},
    \hepth{9712070} ; \\  
    I.\ Kishimoto and N.\ Sasakura, \J{\PL}{B432}{1998}{305},
    \hepth{9712180}.

\bibitem{BB}
    A.\ Hanany and A.\ Zaffaroni,
      \J{J.\ High Energy
      Phys.}{05}{1998}{001}, \hepth{9801134}; \\
    E.\ G. Gimon and M.\ Gremm, 
     \J{\PL}{B433}{1998}{318}, \hepth{9803033} ; \\
    L.\ Randall, Y.\ Shirman and R. von Unge,
     \J{\PR}{D58}{1998}{105005},   {\tt hep-th/} {\tt 9806092} ; \\
    R.\ G.\ Leigh and M.\ Rozali, 
     \J{\PR}{D59}{1999}{026004},  \hepth{9807082}.

\bibitem{CM}
    C.\ G.\ Callan Jr.\ and J.\ M.\ Maldacena,
     \J{\NP}{B513}{1998}{198}, \hepth{9708147}.

\bibitem{Gibb}
    G.\ W.\ Gibbons,
    \J{\NP}{B514}{1998}{603},  \hepth{9709027}.

\bibitem{Hashi}
    A.\ Hashimoto,
    \J{\PR}{D57}{1998}{6441}, \hepth{9711097}.

\bibitem{BER} 
    O.\ Bergman, 
     \J{\NP}{B525}{1998}{104}, \hepth{9712211} ; \\
    O.\ Bergman and B.\ Kol, \J{\NP}{B536}{1998}{149},
    \hepth{9804160}. 
    
\bibitem{Dia}
    D.\ -E.\ Diaconescu, 
     \J{\NP}{B503}{1997}{220}, \hepth{9608163}.

\bibitem{nahm}
    W.\ Nahm, \J{\PL}{B90}{1980}{413}.

\bibitem{Tseytlin}
    A.\ A.\ Tseytlin, 
     \J{\NP}{B501}{1997}{41}, \hepth{9701125}.

\bibitem{Wit}
    E.\ Witten, 
      \J{\NP}{B460}{1996}{335}, \hepth{9510135}.

\bibitem{Nil}
    M.\ F.\ Sohnius, \J{\PR}{128}{1985}{39}.

\bibitem{Str}
    A.\ Strominger,
    \J{\PL}{B383}{1996}{44}, \hepth{9512059}.

\bibitem{BC}
    D.\ Tsimpis, 
     \hepth{9804081}.

\bibitem{BPS}
    E.\ B.\ Bogomol'nyi, \J{Yad.\ Fiz.\ }{24}{1976}{861} ; \\ 
    M.\ K.\ Prasad and C.\ M.\ Sommerfield, \J{\PRL}{35}{1975}{760}.

\bibitem{mono}
    S.\ A.\ Brown, H.\ Panagopoulos and M.\ K.\ Prasad, 
     \J{\PR}{D26}{1982}{854} ; \\
    H.\ Panagopoulos, \J{\PR}{D28}{1983}{380}.

\bibitem{ax}
    L.\ O'Raifeartaigh, S.\ Rouhani and L.\ P.\ Singh,
    \J{\NP}{B206}{1982}{137}.

\bibitem{Bre}
    D.\ Brecher, \J{\PL}{B442}{1998}{117}, 
    \hepth{9804180}.

\end{thebibliography}
\end{document}